\UseRawInputEncoding
\documentclass[superscriptaddress, notitlepage, reprint]{revtex4-1}
\usepackage[english]{babel}
\usepackage{amsmath,amsthm}
\usepackage{amsfonts}
\usepackage[pdfborder={0 0 0}, colorlinks=true, urlcolor=blue, linkcolor=blue, citecolor=blue]{hyperref}
\usepackage{color}
\usepackage{graphicx}
\usepackage{float}
\usepackage{subfigure}
\usepackage{lipsum}
\usepackage{epstopdf}
\usepackage{dcolumn}
\begin{document}

\title{Quantum metrology with quantum Wheatstone bridge composed of Bose systems}
\author{Dong  Xie}\email{xiedong@mail.ustc.edu.cn}
\affiliation{School of Science, Guilin University of Aerospace Technology, Guilin, Guangxi 541004, People's Republic of China}
\author{Chunling Xu}\email{xuchunling@guat.edu.cn}
\affiliation{School of Science, Guilin University of Aerospace Technology, Guilin, Guangxi 541004, People's Republic of China}
\author{Xiwei Yao}
\affiliation{School of Science, Guilin University of Aerospace Technology, Guilin, Guangxi 541004, People's Republic of China}
\author{An Min Wang}
\affiliation{Department of Modern Physics, University of Science and Technology of China, Hefei, Anhui 230026, People's Republic of China}

\begin{abstract}
 The quantum version of a special classical Wheatstone bridge built with a boundary-driven spin system has recently been proposed. We propose a quantum Wheatstone bridge consisting of Bose systems, which can simulate the general classical Wheatstone bridge. Unknown
 coupling can be obtained when the quantum Wheatstone bridge is balanced, which can be determined simply by the homodyne detection. When the expectation value of the homodyne detection is 0, the quantum Wheatstone bridge is unbalanced. Regulate a known coupling strength to make the expectation value of the homodyne detection be proportional to the square root of the initial number of bosons, which means that the quantum Wheatstone bridge is balanced. By calculating the quantum Fisher information, we show that the measurement precision is optimal when the quantum Wheatstone bridge is balanced. And the homodyne detection is close to the optimal measurement in the case of low-temperature baths.

\end{abstract}
\maketitle

\section{Introduction}
\begin{figure}[h]
\includegraphics[scale=0.4]{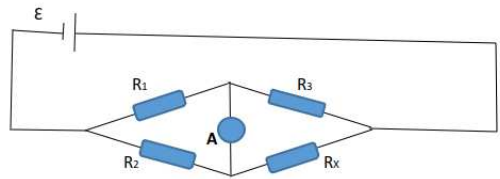}
 \caption{\label{fig.1}Schematic diagram of the classical Wheatstone bridge. It is composed of three controllable resistances ($R_1,\ R_2,\ R_3$), and the Resistance $R_x$ to be measured.  $A$ denotes a galvanometer.}
\end{figure}

The classical Wheatstone bridge is an instrument for accurate measurement of electrical resistance. Although its structure is simple, its accuracy and sensitivity are relatively high, and it has been widely used in medical diagnosis, testing instruments and automatic centering control~\cite{lab1,lab2,lab3}. It was invented by English inventor Christie in 1833~\cite{lab4}, but it became known because Wheatstone~\cite{lab5} was the first to use it to measure resistance. As shown in Fig.~\ref{fig.1}, there are three known and tunable resistances ($R_1,\ R_2,\ R_3$), and the Resistance $R_x$ to be measured. By varying the three tunable resistances until the indicator of galvanometer $A$ is 0, the classical Wheatstone bridge is in balance. At the balance point, the value of $R_x$ can be obtained
\begin{align}
R_x=\frac{R_2}{R_1}R_3.
\label{eq:1}
\end{align}

With the study of smaller and smaller scale, quantum effects are gradually prominent. Based on the quantum estimation theory~\cite{lab6}, there are many works exploring the use of quantum resources to improve parameter measurement precision, such as, quantum magnetometer~\cite{lab7,lab8,lab9}, quantum thermometry~\cite{lab10,lab11,lab12,lab13,lab14,lab15}, quantum interferometry~\cite{lab16}, and Quantum illumination~\cite{lab17,lab18,lab19}.

\begin{figure}[h]
\includegraphics[scale=0.35]{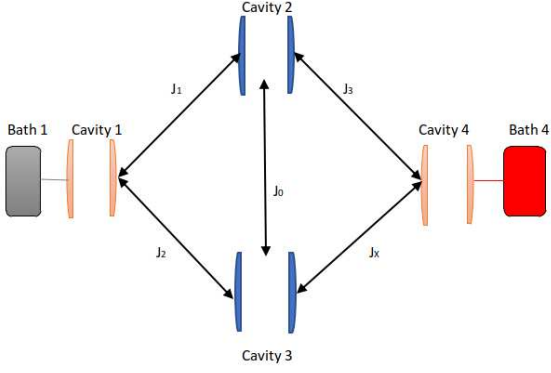}
 \caption{\label{fig.2}Schematic diagram of the quantum Wheatstone bridge. This bridge is composed of 4 cavity modes. There are four tunable and known coupling strengths $J_1,\ J_2,\ J_3, J_0$ and one coupling strength $J_x$ to be measured. Cavity modes 1 and 4 are interacting with two thermal baths with temperature $T_1$ and $T_4$.}
\end{figure}

Recently, Poulsen \textit{et al.}\cite{lab20} utilized a few-body boundary-driven spin chain to build a quantum version of the special
Wheatstone bridge with $R_1=R_2$. In this article, we propose a quantum Wheatstone bridge consisting of Bose systems, which can simulate the general classical Wheatstone bridge, as shown in Fig.~\ref{fig.2}.
Similar to the classical Wheatstone bridge, the unknown coupling strength $J_x$ in the quantum Wheatstone bridge composed of Bose systems can be obtained at the balance point
\begin{align}
J_x=\frac{J_2}{J_1}J_3.
\label{eq:2}
\end{align}
In Ref.~\cite{lab20}, they only considered the case of $J_2=J_1$. And they got the unknown strength at the balance point by the formula $J_x=\lambda J_3$, where the coefficient $\lambda$ was not equal to 1. It deviates from the classical Wheatstone bridge and is therefore not a strictly quantum Wheatstone bridge.

In this article, we are able to simply judge the bridge balance by the local homodyne detection of cavity mode 2 or 3.
When $J_1=J_2$, the quantum Wheatstone bridge can be further simplified with $J_0=0$. By analytically calculating the quantum Fisher information, we show that the homodyne detection is close to the optimal measurement in the case of low-temperature baths.
The quantum fluctuation from the thermal baths will reduce the measurement precision. With the optimal measurement, measurement precision becomes more and more independent with the increase of fluctuation intensity. For resisting the dissipation of cavity modes 2 and 3, the gain process is proposed to maintain the criteria for quantum Wheatstone bridge  balance.


This article is organized as follows. In Section II, we introduce the setup of the quantum Wheatstone bridge. In section III, the  balance criterion of quantum Wheatstone bridge is given. The measurement precision with the homodyne detection is achieved in section IV. By calculating the quantum Fisher information, the optimal precision is obtained in section V. We propose a way to deal with the extra dissipation in section VI. The experiment feasibility of our scheme is discussed in section VII. Finally, we make a simple conclusion in section VIII.
\section{ setup consisting of the cavity modes }
We propose to build the quantum Wheatstone bridge composed of Bose systems, taking the cavity modes as an example, as shown in Fig.~\ref{fig.2}.

The total Hamiltonian can be described by  ($\hbar=\kappa_B=1$ throughout this article )
\begin{align}
H&=\sum_{i=1}^4\omega_ia^\dagger_i a_i+J_1a_1^\dagger a_2+J_2a_1^\dagger a_3+\nonumber\\
&J_3a_4^\dagger a_2+J_xa_4^\dagger a_3+J_0a_2^\dagger a_3+h.c.,
\label{eq:3}
\end{align}
where $a_i$ ($a_i^\dagger$) denotes annihilation (creation) operator of the $i$th cavity mode with the corresponding frequency $\omega_i$; $J_1$, $J_2$, $J_3$, $J_x$, $J_0$ represent the coupling strength between cavity modes 1 and 2, 1 and 3, 2 and 4, 3 and 4, 2 and 3, respectively. The coupling strengths $J_1,\ J_2,\ J_3, J_0$ are known and tunable. The unknown coupling strength $J_x$ is what we want to measure.

The cavity modes 1 and 4 are coupled with two Markovian thermal baths. The whole dynamic can be described by
quantum Langevin-Heisenberg equation
\begin{align}
\dot{\vec{a}}=\mathbf{M}\vec{a}+\vec{a}_{\textmd{in}}
\label{eq:4}
\end{align}
where the cavity mode vector $\vec{a}=(a_1,a_2,a_3,a_4)^\top$, the evolution matrix $\mathbf{M}$ is expressed as
\[
 \mathbf{M}= \left(
\begin{array}{ll}
-i \omega_1-\kappa_1\ \ -i J_1\ \ -iJ_2\ \ \ \ \ \ 0\\
-i J_1\ \ \ \ \ \ \ \ -i \omega_2\ \ -iJ_0\ \ -iJ_3\\
-i J_2\ \ \ \ \ \ \ \ -iJ_0\ \ -i \omega_3\ \ -i J_x\\
\ \ 0\ \ \ \ \ \ \ \ \ \ \ -i J_3\ \ -i J_x\ \ -i \omega_4-\kappa_4
  \end{array}
\right ).
\]
And $\vec{a}_{\textmd{in}}=(\sqrt{2\kappa_1}a_{1\textmd{in}}(t),0,0,\sqrt{2\kappa_4}a_{4\textmd{in}}(t))^T$, where
the input noise operators $a_{j\textmd{in}}$ satisfy that
\begin{align}
\langle a_{j\textmd{in}}^\dagger(t)\rangle=0,\\
\langle a_{j\textmd{in}}^\dagger(t) a_{j\textmd{in}}(t')\rangle=N_j\delta(t-t'),\\
\langle a_{j\textmd{in}} (t')a_{j\textmd{in}}^\dagger(t)\rangle=(N_j+1)\delta(t-t'),
\label{eq:A1}
\end{align}
where the thermal average boson number $N_j=\frac{1}{\exp(\omega_j/T_j)-1}$ with $j=1,4$. $\kappa_j$ is the coupling strength between the cavity mode $j$ and the thermal bath $j$ with the temperature $T_j$. $\langle \mathbf{.}\rangle$ represents the mean over the state of the bath degrees of freedom.
\section{ Quantum Wheatstone bridge balance criterion }
When  $J_x=\frac{J_2}{J_1}J_3$ at the balance point, the Hamiltonian in Eq.~(\ref{eq:3}) can be rewritten with bright mode $A_+=J_1a_2+J_2a_3$ and dark mode $A_-=J_2a_2- J_1a_3$
\begin{align}
H&=\omega_1a^\dagger_1 a_1+\omega_4a^\dagger_4 a_4+\lambda_+A_+^\dagger A_++\lambda_-A_-^\dagger A_-\nonumber\\
&a_1^\dagger A_++J_3/J_1a_4^\dagger A_++h.c.,
\label{eq:A01}
\end{align}
The dark mode $A_-$ decouples with the cavity modes $a_1$ and $a_4$, which are subjected to Markovian thermal baths.
In order to make the Hamiltonian in the above equation be equal to the Hamiltonian in Eq.~(\ref{eq:3}). The values of $\lambda_+$ and $\lambda_-$ should satisfy the following conditions
\begin{align}
\lambda_+J_1^2+\lambda_-J_2^2=\omega_2,\\
\lambda_+J_2^2+\lambda_-J_1^2=\omega_3,\\
\lambda_+-\lambda_-=\frac{J_0}{J_1J_2}.
\label{eq:A2}
\end{align}
The above equations have a solution only if
\begin{align}
\omega_3-\omega_2=J_0(\frac{J_2}{J_1}-\frac{J_1}{J_2}).
\label{eq:12}
\end{align}

When Eq.~(\ref{eq:2}) and Eq.~(\ref{eq:12}) hold, we can get the dark mode $A_-(t)=J_2a_2(t)- J_1a_3(t)$, i.e.,
\begin{align}
A_-(t\rightarrow\infty)=e^{i \varphi(t)}A_-(t=0),
\label{eq:A2}
\end{align}
where the phase $\varphi(t)$ is real. Based on the existence of the dark mode, we next find the detail balance criterion of quantum Wheatstone  bridge.

For $\kappa_1,\kappa_4 \gg|\omega_i-\omega_j|$ with $\{i,j\}=\{1,2,3,4\}$, we can analytically obtain the solution of Eq.~(\ref{eq:4}) (see the detail in Appendix).
The solutions of the modes $a_2(t)$ and $a_3(t)$ can be given by
\[
 \left(
\begin{array}{ll}
a_2(t)\\
a_3(t)
  \end{array}
\right )=e^{\mathbf{M}_2t}\left(
\begin{array}{ll}
{a_2(0)}\\
{a_3(0)}
  \end{array}
\right )+\int_0^te^{\mathbf{M}_2(t-t')}\left(
\begin{array}{ll}
{A_{2\textmd{in}}(t')}\\
{A_{3\textmd{in}}(t')}
  \end{array}
\right ),\tag{14}
\]
where the evolution matrix $\mathbf{M}_2$ is described by
\[
 \mathbf{M}_2= \left(
\begin{array}{ll}
-i\omega_2-\frac{J_1^2}{\kappa_1}-\frac{J_3^2}{\kappa_4}\ \ \ \ \ \ \ -iJ_0-\frac{J_1J_2}{\kappa_1}-\frac{J_3J_x}{\kappa_4}\\
-iJ_0-\frac{J_1J_2}{\kappa_1}-\frac{J_3J_x}{\kappa_4} \ \ -i\omega_3-\frac{J_2^2}{\kappa_1}-\frac{J_x^2}{\kappa_4}
  \end{array}
\right ),\tag{15}
\]
and the noise operators are
\begin{align}
A_{2\textmd{in}}(t)=-\sqrt{2}i[\frac{J_1a_{1\textmd{in}}(t)}{\sqrt{\kappa_1}}
+\frac{J_3a_{4\textmd{in}}(t)}{\sqrt{\kappa_4}}],\tag{16}\\
A_{3\textmd{in}}(t)=-\sqrt{2}i[\frac{J_2a_{1\textmd{in}}(t)}{\sqrt{\kappa_1}}
+\frac{J_xa_{4\textmd{in}}(t)}{\sqrt{\kappa_4}}].\tag{17}
\end{align}
When $J_1\neq J _2$ and the quantum Wheatstone bridge is balanced,  the eigenvalues of $\mathbf{M}_2$ can be achieved
\begin{align}
E_1&=-\frac{i(J_1^2\omega_2-J_2^2\omega_3)}{J_1^2-J_2^2},\tag{18}\label{eq:18}\\
E_2&=-\frac{(J_1^2+J_2^2)(J_3^2\kappa_1+J_1^2\kappa_4)}{J_1^2\kappa_1\kappa_4}-\frac{i(J_1^2\omega_2-J_2^2\omega_3)}{J_1^2-J_2^2}.\tag{19}
\end{align}
The eigenvalue $E_1$ is pure imaginary, which means that $ e^{\mathbf{M}_2t}\neq\mathbf{0}$ for the long time limit $t\rightarrow\infty$. It leads to that the expectations of the cavity modes 2 and 3 are described by
\begin{align}
\langle a_2(t\rightarrow\infty)\rangle=\frac{e^{E_1t}J_2}{J_1^2+J_2^2}\langle J_2 a_2(0)-J_1a_3(0)\rangle;\tag{20}\label{20}\\
\langle a_3(t\rightarrow\infty)\rangle=\frac{e^{E_1t}J_1}{J_1^2+J_2^2}\langle J_1a_3(0)-J_2 a_2(0)\rangle,\tag{21}\label{21}
\end{align}
where $\langle \mathbf{\star}\rangle$ denotes the mean over the initial state of the cavity modes.
When $J_x\neq\frac{J_2}{J_1}J_3$, the real parts of the eigenvalues of $\mathbf{M}_2$ are all negative. As a result, the expectation values $\langle a_2(t\rightarrow\infty)\rangle=\langle a_3(t\rightarrow\infty)\rangle=0$.

When $J_1=J _2$, the frequency $\omega_2$ must be equal to $\omega_3$ for achieving the balanced quantum Wheatstone bridge. In this case, the coupling strength $J_0$ can be arbitrary value, which is different from the case of $J_1\neq J _2$. The corresponding eigenvalues of $\mathbf{M}_2$ are
\begin{align}
E'_1&=i(J_0-\omega_3),\tag{22}\\
E'_2&=i(J_0-\omega_3)-\frac{2J_1^2}{\kappa_1}-\frac{2J_3^2}{\kappa_4}.\tag{23}
\end{align}
For the long time limit $t\rightarrow\infty$, we get the similar form as the previous results
\begin{align}
\langle a_2(t\rightarrow\infty)\rangle=\frac{e^{E'_1t}}{2}\langle a_2(0)-a_3(0)\rangle;\tag{24}\\
\langle a_3(t\rightarrow\infty)\rangle=\frac{e^{E'_1t}}{2}\langle a_3(0)- a_2(0)\rangle.\tag{25}
\end{align}

From Eq.~(\ref{eq:18}), we can derive that ${\lim}_{J_1\rightarrow J_2}E_1=-i\omega_3$, which is different from $E'_1$ unless that $J_0$ is equal to 0.
As a result, there is a discontinuity for the phase of $\langle a_2(t\rightarrow\infty)\rangle$ and $\langle a_3(t\rightarrow\infty)\rangle$.

The detail balance criterion of quantum Wheatstone  bridge can be be summarized as follows: With the initial state of cavity modes 2 and 3 in the coherent state $|\alpha\rangle|0\rangle$, it is balanced when the expectation values of $\langle a_2(t\rightarrow\infty)\rangle$ and $\langle a_3(t\rightarrow\infty)\rangle$  are not equal to 0; otherwise it is not balanced.

\section{ local homodyne detection}
By the homodyne detection of the cavity mode $a_2(t\rightarrow\infty)$ or $a_3(t\rightarrow\infty)$, we can obtain the information of $J_x$. The uncertainty of $J_x$ can be achieved by the error transfer formula
\begin{align}
\delta J_x=\frac{\sqrt{\langle X^2\rangle-\langle X\rangle^2}}{|\frac{d\langle X\rangle}{dJ_x}|},\tag{26}\label{eq:26}
\end{align}
where $X$ denotes the measurement operator.

\subsection{ In the case of $J_1=J_2$}

When $J_1=J_2$ and $\omega_2=\omega_3$, we can analytically obtain the differential coefficients
\begin{align}
\lim_{J_x\rightarrow J_3}\frac{d}{dJ_x}\langle a_2(t\rightarrow\infty)\rangle=\frac{J_3\kappa_1e^{it(J_0-\omega_3)}\langle a_2(0)\rangle}{2(J_3^2\kappa_1+J_2^2\kappa_4+iJ_0\kappa_1\kappa_4)},\tag{27}\\
\lim_{J_x\rightarrow J_3}\frac{d}{dJ_x}\langle a_3(t\rightarrow\infty)\rangle=\frac{-J_3\kappa_1e^{it(J_0-\omega_3)}\langle a_3(0)\rangle}{2(J_3^2\kappa_1+J_2^2\kappa_4+iJ_0\kappa_1\kappa_4)}.\tag{28}
\end{align}

For the long evolution time $t>\tau=1/( \frac{2J_1^2}{\kappa_1}+\frac{2J_3^2}{\kappa_4})$, we can derive the taylor series expansion

\begin{align}
\langle a_2(t)\rangle=&\frac{1}{2}e^{E'_1t-\Gamma y^2 t-o(y^3)t}\langle (1+\Lambda y+o(y^2)) a_2(0)\nonumber\\
&-(1+o(y^2))a_3(0)\rangle;\tag{29}\\
\langle a_3(t)\rangle=&-\frac{1}{2}e^{E'_1t-\Gamma y^2 t-o(y^3)t}\langle 1+o(y^2) a_2(0)\nonumber\\
&-(1-\Lambda y+o(y^2))a_3(0)\rangle,\tag{30}
\end{align}
where the coefficients $\Gamma=\frac{J_2^2+iJ_0\kappa_1}{2(J_3^2\kappa_1+J_2^2\kappa_4+iJ_0\kappa_1\kappa_4)}$, $\Lambda=\frac{J_3\kappa_1}{J_3^2\kappa_1+J_2^2\kappa_4+iJ_0\kappa_1\kappa_4}$ and the difference $y=J_x-J_3$. For larger difference, $y\geq1$, the expectation values of $\langle a_2(t)\rangle$ and $\langle a_3(t)\rangle$ go to 0 very quickly. Therefore, we consider that $y<1$. It leads to that higher-order terms $o(y^3)t$ and $o(y^2)$ can be ignored.

We consider the homodyne  detection with the local quadrature operator $X_{\varphi_j}=a_je^{-i\varphi_j}+a_j^\dagger e^{-i\varphi_j}$, where $\varphi_j$ represents phase with $j=2,3$.
The initial states of cavity 2 and 3 are given by the classical coherent state $|\alpha\rangle_2|0\rangle_3$, with the real value $\alpha$.
By utilizing Eq.~(\ref{eq:26}), we can obtain the measurement precision of $J_x$ with the quadrature operator $X_{\varphi_2}$
\begin{align}
\delta J_x=\frac{{\sqrt{1+f}}}{|Re[(\Lambda-2\Gamma y t(1+\Lambda y))e^{i\varphi_2+E'_1t-\Gamma y^2 t}]\alpha|},\tag{31}
\end{align}
where the quantum fluctuation $f=\frac{J_1^2N_1\kappa_4+J_3^2N_4\kappa_1}{J_1^2\kappa_4+J_3^2\kappa_1}$.
From the above equation, one can find that $\delta J_x$ takes the minimum value at $y=0$. It shows that the optimal precision of $J_x$ can be obtained when the Wheatstone bridge is balanced:
\begin{align}
\delta J_x(y=0)=\frac{\sqrt{1+f}{\sqrt{(J_3^2\kappa_1+J_2^2\kappa_4)^2+J_0^2\kappa_1^2\kappa_4^2 }}}{J_3\kappa_1|\cos(\varphi_2+\theta+(J_0-\omega_3)t)|\alpha},\tag{32}
\end{align}
where the phase $\theta$ satisfies that $e^{i\theta}=\Lambda/|\Lambda|$.
When the phase satisfies that $\varphi_2+\theta+(J_0-\omega_3)t=n \pi $( $n$ represents an integer), the optimal precision by the homodyne detection can be expressed as
\begin{align}
\delta J^o_x=\frac{\sqrt{1+f}\sqrt{(J_3^2\kappa_1+J_2^2\kappa_4)^2+J_0^2\kappa_1^2\kappa_4^2}}{J_3\kappa_1\alpha}.\tag{33}\label{33}
\end{align}

Let $J_0=0$, the precision can be further improved,
\begin{align}
\delta J^o_x(J_0=0)=\frac{\sqrt{1+f}(J_3^2\kappa_1+J_2^2\kappa_4)}{J_3\kappa_1\alpha}.\tag{34}
\end{align}

\subsection{ In the case of $J_1\neq J_2$}
In the case of $J_1\neq J_2$, we can get the derivative at the balance point

\begin{align}
\lim_{J_x\rightarrow J_3}\frac{d}{dJ_x}\langle a_2(t\rightarrow\infty)\rangle=\frac{2J_1^3J_2J_3\kappa_1e^{E_1t}\langle a_2(0)\rangle}{\mu}\nonumber\\
-\frac{(J_1^2-J_2^2)J_1^2J_3\kappa_1e^{E_1t}\langle a_3(0)\rangle}{\mu},\tag{35}\label{35}\\
\lim_{J_x\rightarrow J_3}\frac{d}{dJ_x}\langle a_3(t\rightarrow\infty)\rangle=-\frac{2J_1^3J_2J_3\kappa_1e^{E_1t}\langle a_3(0)\rangle}{\mu}\nonumber\\
-\frac{(J_1^2-J_2^2)J_1^2J_3\kappa_1e^{E_1t}\langle a_2(0)\rangle}{\mu},\tag{36}\label{36}
\end{align}
where $\mu=(J_1^2+J_2^2)^2(J_3^2\kappa_1+J_1^2\kappa_4+i J_0J_1\kappa_1\kappa_4/J_2)$.
The measurement precision of $J_x$ can be given by
\begin{align}
\delta J_x=\frac{\sqrt{1+f_c}|\mu|}{4J_1^3J_2J_3\kappa_1\alpha|\cos(\varphi_2+\theta'+\frac{J_1^2\omega_2-J_2^2\omega_3}{J_1^2-J_2^2}t)|},\tag{37}
\end{align}
where  the phase $\theta'$ satisfies that $e^{i\theta'}=|\mu|/\mu$ and the quantum fluctuation $f_c=\frac{2J_1^2N_1\kappa_4+2J_3^2N_4\kappa_1}{(J_1^2\kappa_4+J_3^2\kappa_1)(J_1^2+J_2^2)}$.
When the phase satisfies that $\varphi_2+\theta'-iE_1 t=n \pi $, the optimal precision with the homodyne detection can be expressed as
\begin{align}
\delta J^o_x=\frac{\sqrt{1+f_c}|\mu|}{4J_1^3J_2J_3\kappa_1\alpha}.\tag{38}\label{38}
\end{align}

\begin{figure}[h]
\includegraphics[scale=0.6]{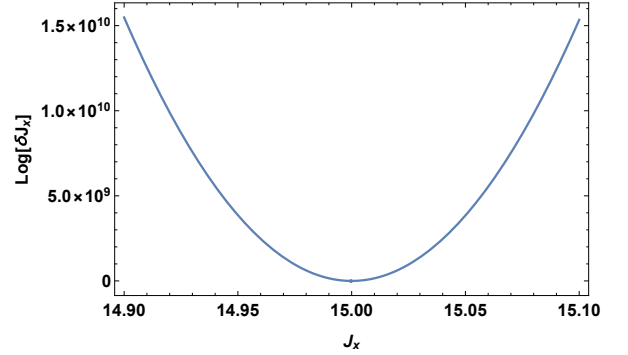}
 \caption{\label{fig.3}A plot of the logarithm of the uncertainty of $J_x$ versus $J_x$. Here, the dimensionless parameters are set as: $\omega_2=100$, $\omega_3=101$, $\alpha=10000$, $\kappa_1=10$, $\kappa_4=10$, $J_1=10$,$J_2=15$,$J_3=10$, and $J_0=(\omega_3-\omega_2)J_1J_2/(J_2^2-J_1^2)=1.2$. At the balance point, the unknown coupling strength $J_x=J_2J_3/J_1=15$.}
\end{figure}

When $J_1=J_2$, Eq.~(\ref{38}) becomes Eq.~(\ref{33}). It shows that the optimal precision is still continuous at the point $J_1=J_2$ and not affected by the discontinuity for the phase of $\langle a_2(t\rightarrow\infty)\rangle$ and $\langle a_3(t\rightarrow\infty)\rangle$. This is mainly because  the measurement angle $\varphi_2$ can compensate of the effect of phase.

As shown in Fig.~\ref{fig.3}, we can see that the minimum uncertainty of $J_x$ appears when the quantum Wheatstone bridge is balanced, i.e., $J_x=J_2J_3/J_1=15$. Consistent with the previous results, the measurement precision at the balance point is optimal.

\section{Quantum Fisher information of $J_x$}

The quantum Cram\'{e}r-Rao bound \cite{lab21,lab22,lab23,lab24} provides the bound on the estimation precision of parameter $J_x$,
\begin{align}
\delta J_x\geq\frac{1}{\mathcal{F}[J_x]},\tag{39}
\end{align}
where $\mathcal{F}[J_x]$ denotes the quantum Fisher information of $J_x$, which is encoded into the cavity modes.
Due to that Hamiltonian is quadratic and the initial state is Gaussian, the steady state of cavity modes 2 and 3 is Gaussian too \cite{lab24a}. The Quantum Fisher information for Gaussian state is obtained through the fidelity by Pinel \textit{et al.} in 2013 \cite{lab25},
\begin{align}
\mathcal{F}[J_x]=\frac{1}{2(1+P_{J_x}^2)}\textmd{Tr}[(\mathcal{C}^{-1}_{J_x}\mathcal{C}^{'}_{J_x})^2]+\frac{2P_{J_x}'^2}{1-P_{J_x}^4}\nonumber\\
+{\langle\mathbf{X}^\top\rangle}'_{J_x}\mathcal{C}^{-1}_{J_x}\langle\mathbf{X}\rangle'_{J_x},\tag{40}
\label{eq:40}
\end{align}
where $P_{J_x}=\frac{1}{2 d}$, $d=\sqrt{\textmd{Det }\mathcal{C}_{J_x}}$ and $A'_{J_x}$ is the term by term derivative of $A_{J_x}$ with respect to ${J_x}$.
The entries of the covariance matrix $\mathcal{C}$ are defined as $\mathcal{C}_{ij}:=\frac{1}{2}\langle \{\mathbf{X}_i,\mathbf{X}_j\}\rangle-\langle \mathbf{X}_i\rangle\langle \mathbf{X}_j\rangle$, with  $\langle\bullet\rangle$ the expectation value.
The vector of quadrature operators is $\mathbf{X}=(q_2,q_3,p_2,p_3)^\top$, where
the quadrature operators are defined as $q_j :=(a_j+a^\dagger_j)$ and $p_j :=\frac{1}{i}(a_j-a^\dagger_j)$.

By utilizing Eq.~(\ref{20}) and Eq.~(\ref{21}), we can obtain the covariance matrix
\[
 \mathbf{\mathcal{C}}= \left(
\begin{array}{ll}
1+f_c\ \ \ \ \frac{f_cJ_2}{J_1}\ \ \ \ \ \ \ \ \ 0\ \ \ \ \ \ \ \ \ \ 0\\
\frac{f_cJ_2}{J_1} \ \ \ \ 1+\frac{f_cJ_2^2}{J_1^2}\ \ \ \ \ \ 0\ \ \ \ \ \ \ \ \ \ 0\\
\ \ 0\ \ \ \ \ \ \ \ \ 0\ \ \ \ \ \ \ \ \ 1+f_c\ \ \ \ \ \frac{f_cJ_2}{J_1}\\
\ \ 0\ \ \ \ \ \ \ \ \ 0\ \ \ \ \ \ \ \ \ \ \frac{f_cJ_2}{J_1}\ \ \ \ \ \ 1+\frac{f_cJ_2^2}{J_1^2}
  \end{array}
\right ).\tag{41}
\]
And using Eq.~(\ref{35}) and Eq.~(\ref{36}), the derivative of the vector of quadrature operators is given by
\begin{align}
\langle\mathbf{X}\rangle'_{J_x}=\frac{\alpha J_1^2J_3\kappa_1}{|\mu|}(4J_1J_2\cos \phi,
2(J_2^2-J_1^2)\cos \phi,\nonumber\\
4J_1J_2\sin \phi,2(J_2^2-J_1^2)\sin\phi),\tag{42}
\end{align}
where $\phi=\theta'-iE_1t$.

When the input probes are sufficiently strong, the quantum Fisher information can be dominated by the last term in Eq.~(\ref{eq:40})\cite{lab26}
\begin{align}
\mathcal{F}[J_x]\approx{\langle\mathbf{X}^\top\rangle}'_{J_x}\mathcal{C}^{-1}_{J_x}\langle\mathbf{X}\rangle'_{J_x}\tag{43}\\
=\frac{(\alpha J_1^2J_3\kappa_1)^2}{|\mu|^2}\frac{4(1+f_c)J_1^2(J_1^2+J_2^2)^2}{(1+f_c)J_1^2+f_cJ_2^2}\tag{44}
\label{eq:44}
\end{align}
According to the quantum Cram\'{e}r-Rao bound, the uncertainty of $J_x$ is
\begin{align}
\delta J_x\geq\frac{g\sqrt{1+f_c}|\mu|}{4J_1^3J_2J_3\kappa_1\alpha},\tag{45}
\end{align}
where the coefficient $g=\frac{2J_2\sqrt{(1+f_c)J_1^2+f_cJ_2^2}}{(J_1^2+J_2^2)(1+f_c)}$.

By a simple calculation, we can prove that
\begin{align}
1-g^2=\frac{[(1+f_c)J_1^2+(f_c-1)J_2^2]^2}{(J_1^2+J_2^2)^2(1+f_c)^2}\geq0.\tag{46}
\end{align}
It means that the coefficient $g\leq1$. When $J_2^2=\frac{1+f_c}{1-f_c} J_1^2$,  the equality holds, i.e., $g=1$. Comparing with Eq.~(\ref{38}), it shows that the homodyne detection is the optimal measurement in the case of $J_2^2=\frac{1+f_c}{1-f_c}J_1^2$ and $\alpha\gg1$. In order to get the equation $J_2^2=\frac{1+f_c}{1-f_c}J_1^2$, the quantum fluctuation $f_c$ must be less than 1. It means that the temperature of both thermal baths should be relatively small. When $f_c=0$, i.e., the temperature of both thermal baths is 0 ($T_1=T_4=0$)
\begin{align}
\delta J_x\geq\frac{|\mu| }{2J_1^2J_3\kappa_1{(J_1^2+J_2^2)}\alpha}.\tag{47}
\end{align}

For large quantum fluctuation $f_c>1$, the homodyne detection is not the optimal measurement. With the homodye detection, the precision of $J_x$ decreases as the quantum fluctuation term $f_c$ increases. With the optimal measurement, the precision of  $J_x$ becomes increasingly independent of the quantum fluctuation term $f_c$. For $f_c\gg1$, the precision of  $J_x$ is given by

\begin{align}
\delta J_x\geq\frac{|\mu|}{2J_1^3J_3\kappa_1\sqrt{J_1^2+J_2^2}\alpha}.\tag{48}
\end{align}

Especially, in the case of $J_1=J_2$, $\omega_2=\omega_3$ and $J_0=0$, the optimal measurement precision of $J_x$ is given by
\begin{align}
\delta J_x^o=\frac{\sqrt{1+2f}(J_3^2\kappa_1+J_2^2\kappa_4)}{J_3\kappa_1\alpha\sqrt{1+f}}.\tag{49}
\end{align}
From the above equation, we can see that the measurement uncertainty of $J_x$ increases with the quantum fluctuation $f_c$.
And the ratio $\delta J_x(f_c=0)/\delta J_x(f_c\rightarrow\infty)=1/\sqrt{2}$. It shows that quantum fluctuations can reduce the precision by up to a factor of $1/\sqrt{2}$. However, the uncertainty with the local homodyne detection still increases. Hence, it is necessary to construct an optimal nonlocal measurement operator to reduce the influence of the quantum fluctuation.

\section{ Dissipation of cavity 2 and 3}
When the cavity modes 2 and 3 are not perfect, there are intrinsic losses. This will prevent us from determining whether quantum Wheatstone bridge is balanced. This can be overcome by using gains to make up for losses. The gain process can be achieved by coupling with an amplifying channel \cite{lab26} or  harnessing the feeding fields \cite{lab27,lab28}.

Including the intrinsic losses and the gains, quantum Langevin-Heisenberg equation of the cavity modes 2 and 3 can be described as
\begin{align}
\dot{a_2}=-iJ_1a_1+(-i\omega_2-\kappa_2+\gamma_2)a_2-iJ_0a_3-iJ_3a_4+\nonumber\\
\sqrt{2k_2}a_{2\textmd{in}}(t)-\sqrt{2\gamma_2}d_{2\textmd{in}}^\dagger(t);\tag{50}\\
\dot{a_3}=-iJ_2a_1-iJ_0a_2+(-i\omega_3-\kappa_3+\gamma_3)a_3-iJ_xa_4+\nonumber\\
\sqrt{2k_3}a_{3\textmd{in}}(t)-\sqrt{2\gamma_3}d_{3\textmd{in}}^\dagger(t);\tag{51}
\label{eq:51}
\end{align}
where $\kappa_2$, $\kappa_3$ are the intrinsic loss rates of the cavity modes 2, 3, respectively; $\gamma_2$, $\gamma_3$ are the gain rates of the cavity modes 2,3 respectively; the noise operator $Y_{j\textmd{in}}(t)$ satisfies
\begin{align}
\langle Y_{j\textmd{\textmd{in}}}^\dagger(t)\rangle=0,\tag{52}\\
\langle Y_{j\textmd{\textmd{in}}}^\dagger(t) Y_{j\textmd{in}}(t')\rangle=0,\tag{53}\\
\langle Y_{j\textmd{\textmd{in}}} (t')Y_{j\textmd{in}}^\dagger(t)\rangle=\delta(t-t'),\tag{54}
\label{eq:54}
\end{align}
where $Y={a,d}$ and $j=2,3$.

Let the gain rate $\gamma_j$ be equal to the corresponding loss rate $\kappa_j$($\kappa_j=\gamma_j$), the quantum Wheatstone bridge balance criterion is reestablished. In the case of $\kappa_j=\gamma_j$, the solution can be obtained
\[
 \left(
\begin{array}{ll}
a_2(t)\\
a_3(t)
  \end{array}
\right )=e^{\mathbf{M}_2t}\left(
\begin{array}{ll}
{a_2(0)}\\
{a_3(0)}
  \end{array}
\right )+\int_0^te^{\mathbf{M}_2(t-t')}\left(
\begin{array}{ll}
{A_{2\textmd{in}}(t')}\\
{A_{3\textmd{in}}(t')}
  \end{array}\tag{55}
\right ),
\]
where the noise operators are redefined as
\begin{align}
A_{2\textmd{in}}(t)=-\sqrt{2}i[\frac{J_1a_{1\textmd{in}}(t)}{\sqrt{\kappa_1}}
+\frac{J_3a_{4\textmd{in}}(t)}{\sqrt{\kappa_4}}]+\nonumber\\\sqrt{2k_2}(a_{2\textmd{in}}(t)-d_{2\textmd{in}}^\dagger(t)),\tag{56}\\
A_{3\textmd{in}}(t)=-\sqrt{2}i[\frac{J_2a_{1\textmd{in}}(t)}{\sqrt{\kappa_1}}
+\frac{J_xa_{4\textmd{in}}(t)}{\sqrt{\kappa_4}}]+\nonumber\\\sqrt{2k_3}(a_{3\textmd{in}}(t)-d_{3\textmd{in}}^\dagger(t)).\tag{57}
\end{align}
For simplicity, we assume $\kappa_2=\kappa_3$.
As a result, the optimal measurement precision  can be described as
\begin{align}
\delta J_x\geq\frac{g\sqrt{1+f'_c}|\mu|}{4J_1^3J_2J_3\kappa_1\alpha},\tag{58}
\end{align}
where the quantum fluctuation $f'_c=f_c+\frac{\kappa_1\kappa_2\kappa_4}{2(J_1^2\kappa_4+J_3^2\kappa_1)}$. The increment of quantum fluctuation $f'_c-f_c$ is from the dissipation and the gain environment of the cavity modes 2 and 3.
The gain can compensate for the dissipation and reach the balanced quantum Wheatstone bridge, but it can not completely eliminate the effect of quantum fluctuation from the dissipative cavity modes 2 and 3 on the estimation precision.
For the quantum fluctuation $f_c\gg1$, the estimation precision is independent of $f_c$. In this case, the dissipation and the gain environment of the cavity modes 2 and 3 do not result in a loss of the estimation precision.

\section{Possible experimental implementations}
Our model with $J_0=0$ can be realized in the opto-mechanical system. Let the cavity modes 2 and 3 become two mechanical modes. The total Hamiltonian can be rewritten as
\begin{align}
H_{\textmd{om}}&=\sum_{i=1}^4\omega_ia^\dagger_i a_i+G_1a_1^\dagger a_1(a_2+a_2^\dagger)+G_2a_1^\dagger a_1(a_3+a_3^\dagger)+\nonumber\\
&G_3a_4^\dagger a_4(a_2+a_2^\dagger)+G_xa_4^\dagger a_4(a_3+a_3^\dagger),\tag{59}
\label{eq:59}
\end{align}
where $a_2$ and $a_3$ represent the bosonic annihilation operators of the mechanical modes.
Replacing the cavity modes with the sum of its quantum fluctuation operator classical mean value, the Hamiltonian can be linearized in the interaction picture as \cite{lab29}
\begin{align}
H_{\textmd{om,int}}=J_1a_1^\dagger a_2+J_2a_1^\dagger a_3+J_3a_4^\dagger a_2+J_xa_4^\dagger a_3+h.c.,\tag{60}
\label{eq:60}
\end{align}
where $J_i=G_i\beta_i$ is the effective optomechanical coupling strength and classical amplitude $\beta_i$ can be controlled by
the classical driving with $i=(1,\ 2,\ 3,\ x)$.

In experiment, this setup has been created by a superconducting circuit of aluminum on a
sapphire substrate \cite{lab30}, the microchip circulator device composed of three high-impedance spiral inductors
capacitively coupled to the in-plane vibrational modes of a dielectric nanostring mechanical resonator \cite{lab31}. Therefore, our proposed scheme can be implemented using current experimental technology.
\section{Conclusion}
We utilized Bose system to construct the general quantum Wheatstone bridge. We propose a more feasible criterion for determining the balance of the quantum Wheatstone bridge. By calculating the quantum Fisher information, we show that the local homodyne detection is close to the optimal measurement with low temperature of the Markovian baths. With the optimal measurement, the measurement precision of the unknown coupling strength is independent of the quantum fluctuation from the baths with high temperature. In order to maintain the balance criterion of the quantum Wheatstone bridge, extra gain process is proposed to resist the dissipation. Our proposed scheme can be realized in the current optomechanical systems, which will facilitate the precision measurement of photomechanical coupling strength.

\section*{Acknowledgements}
This research was supported by the National Natural Science Foundation of China (Grant No. 62001134), Guangxi Natural Science Foundation (Grant No.2020GXNSFAA159047, 2020GXNSFAA159007), and National Key R\&D Program of China (Grant No. 2018YFB1601402-2).

\section*{Appendix A}
Using the slowly varying operators $S_i=a_ie^{i\omega_it}$, the Eq.(A2) can be rewritten as
\begin{align}
\dot{\vec{S}}=\mathbf{M}\vec{S}+\vec{a'}_{in}\tag{A1}
\label{A1}
\end{align}
where $\vec{a'}_{in}=(e^{i\omega_1t}\sqrt{2\kappa_1}a_{1in}(t),0,0,e^{i\omega_4t}\sqrt{2\kappa_4}a_{4in}(t))^T$, and the evolution matrix $\mathbf{M}$ is expressed as
\[
 \mathbf{M'}= \left(
\begin{array}{ll}
-\kappa_1\ \ -i J'_1\ \ -iJ'_2\ \ \ \ \ \ 0\\
-i J'^*_1 \ \ \ \ 0\ \ \ \ -iJ'_0\ \ -iJ'_3\\
-i J'^*_2\ \ -iJ'^*_0\ \ \ 0\ \ \ \ -i J'_x\\
\ \ 0\ \ \ \ -i J'^*_3\ \ -i J'^*_x\ \ -\kappa_4
  \end{array}
\right ),\tag{A2}
\]
where $J'_1=J_1e^{-i\Delta_{21}t}$, $J'_2=J_2e^{-i\Delta_{31}t}$, $J'_3=J_3e^{-i\Delta_{42}t}$, $J'_x=J_xe^{-i\Delta_{43}t}$ and $J'_0=J_0e^{-i\Delta_{32}t}$ with the frequency detunings between two cavity modes $\Delta_{ij}=\omega_i-\omega_j$.

We can obtain the formal solutions of $S_1$ and $S_4$
\begin{align}
S_1=-i\int_0^tdt'e^{-\kappa_1(t-t')}[J'_1S_2(t')+J'_2S_3(t')]\nonumber\\
+\int_0^tdt'e^{-\kappa_1(t-t')}[\sqrt{2\kappa_1}a'_{1in}(t')];\tag{A3}\label{A3}\\
S_4=-i\int_0^tdt'e^{-\kappa_4(t-t')}[J'^*_3S_2(t')+J'^*_xS_3(t')]\nonumber\\
+\int_0^tdt'e^{-\kappa_4(t-t')}[\sqrt{2\kappa_4}a'_{4in}(t')].\tag{A4}
\label{A4}
\end{align}
We consider that the dissipation rates of the cavity modes $a_1$ and $a_4$ is much larger than the cases of the cavity modes $a_2$ and $a_3$, i.e., $\kappa_1$,$\kappa_4\gg \kappa_2$,$\kappa_3$, where we assume that the dissipation rates of the cavity modes $a_2$ and $a_3$: $\kappa_2=0$,$\kappa_3=0$ for simplicity. The changes of mode $S_2$ and $S_3$ are small within the range of the integration of the cavity modes $S_1$ and $S_4$. Therefore, we can set $S_2(t')=S_2(t)$ and $S_3(t')=S_3(t)$ in Eq.~(\ref{A3}) and Eq.~(\ref{A4}) to obtain
\begin{align}
S_1=\frac{-iJ_1S_2}{\kappa_1-i\Delta_{21}}e^{-i\Delta_{21}t}+\frac{-iJ_2S_3}{\kappa_1-i\Delta_{31}}e^{-i\Delta_{31}t}\nonumber\\
+\frac{\sqrt{2\kappa_1}a_{1\textmd{in}}}{\kappa_1}e^{-i\omega_1t};\tag{A5}\\
S_4=\frac{-iJ_3S_2}{\kappa_4-i\Delta_{24}}e^{-i\Delta_{24}t}+\frac{-iJ_xS_3}{\kappa_4-i\Delta_{34}}e^{-i\Delta_{34}t}\nonumber\\
+\frac{\sqrt{2\kappa_4}a_{4\textmd{in}}}{\kappa_4}e^{-i\omega_4t}.\tag{A6}
\label{eq:A6}
\end{align}
Substituting the above equations into Eq.~(\ref{A1}), the modes $S_1$ and $S_4$ can be adiabatically eliminated. The corresponding evolution equations of the
 modes $S_2$ and $S_3$ are reduced to
\begin{align}
\dot{S_2}=(-\frac{J_1^2}{\kappa_1-i\Delta_{21}}-\frac{J_3^2}{\kappa_1-i\Delta_{24}})S_2+\nonumber\\ (-\frac{J_1J_2}{\kappa_1-i\Delta_{31}}-\frac{J_3J_x}{\kappa_4-i\Delta_{34}}-iJ_0)e^{-i\Delta_{32}t}S_3-\nonumber\\
(\frac{J_1a_{1\textmd{in}}}{\sqrt{\kappa_1}}\tag{A7}
+\frac{J_3a_{4\textmd{in}}}{\sqrt{\kappa_4}})\sqrt{2}ie^{-i\omega_2t};\\
\dot{S_3}=(-\frac{J_1J_2}{\kappa_1-i\Delta_{21}}-\frac{J_3J_x}{\kappa_1-i\Delta_{24}}-iJ_0)e^{-i\Delta_{23}t}S_2+\nonumber\\ (\frac{J_2^2}{\kappa_1-i\Delta_{31}}-\frac{J_x^2}{\kappa_4-i\Delta_{34}})S_3-\nonumber\\
(\frac{J_2a_{1\textmd{in}}}{\sqrt{\kappa_1}}
+\frac{J_xa_{4\textmd{in}}}{\sqrt{\kappa_4}})\sqrt{2}ie^{-i\omega_3t}.\tag{A8}
 \end{align}

For $\kappa_1,\kappa_4 \gg\Delta_{ij}$, combine these equations with $S_2=a_2e^{i\omega_2t}$ and $S_3=a_3e^{i\omega_3t}$, we can obtain
\[
 \left(
\begin{array}{ll}
\dot{a}_2(t)\\
\dot{a}_3(t)
  \end{array}
\right )=\mathbf{M}_2\left(
\begin{array}{ll}
{a_2(t)}\\
{a_3(t)}
  \end{array}
\right )+\left(
\begin{array}{ll}
{A_{2\textmd{in}}(t)}\\
{A_{3\textmd{in}}(t)}
  \end{array}
\right ),\tag{A9}
\]
where the evolution matrix $\mathbf{M}_2$ is described by
\[
 \mathbf{M}_2= \left(
\begin{array}{ll}
-i\omega_2-\frac{J_1^2}{\kappa_1}-\frac{J_3^2}{\kappa_4}\ \ \ \ \ \ \ -iJ_0-\frac{J_1J_2}{\kappa_1}-\frac{J_3J_x}{\kappa_4}\\
-iJ_0-\frac{J_1J_2}{\kappa_1}-\frac{J_3J_x}{\kappa_4} \ \ -i\omega_3-\frac{J_2^2}{\kappa_1}-\frac{J_x^2}{\kappa_4}
  \end{array}
\right ),\tag{A10}
\]
and the noise operators are
\begin{align}
A_{2in}(t)=-\sqrt{2}i[\frac{J_1a_{1\textmd{in}}(t)}{\sqrt{\kappa_1}}
+\frac{J_3a_{4\textmd{in}}(t)}{\sqrt{\kappa_4}}],\tag{A11}\\
A_{3in}(t)=-\sqrt{2}i[\frac{J_2a_{1\textmd{in}}(t)}{\sqrt{\kappa_1}}
+\frac{J_xa_{4\textmd{in}}(t)}{\sqrt{\kappa_4}}].\tag{A12}
\end{align}
The solutions of the modes $a_2(t)$ and $a_3(t)$ can be obtained
\[
 \left(
\begin{array}{ll}
a_2(t)\\
a_3(t)
  \end{array}
\right )=e^{\mathbf{M}_2t}\left(
\begin{array}{ll}
{a_2(0)}\\
{a_3(0)}
  \end{array}
\right )+\int_0^te^{\mathbf{M}_2(t-t')}\left(
\begin{array}{ll}
{A_{2\textmd{in}}(t')}\\
{A_{3\textmd{in}}(t')}
  \end{array}
\right ).\tag{A13}
\]


\begin{thebibliography}{9}

\vspace{3mm}
 \bibitem{lab1}El Mehdi Boujamaa, Yannick Soulie, Fr\'{e}d\'{e}rick Mailly, Laurent Latorre, Pascal Nouet, Rejection of Power Supply Noise in Wheatstone Bridges : Application to Piezoresistive MEMS,  Dans Symposium on Design, Test, Integration and Packaging of MEMS/MOEMS - DTIP 2008, Nice : France (2008).
 \bibitem{lab2}E. Halvorsen, S. Husa, Bridge configurations in piezoresistive two-axis accelerometers, Dans Symposium on Design, Test, Integration and Packaging of MEMS/MOEMS - DTIP 2007, Stresa, lago Maggiore : Italie (2007).
 \bibitem{lab3}S. H. Mennema, J. H. T. Ransley, G. Burnell, J. L. MacManus-Driscoll, E. J. Tarte, and M. G. Blamire, Normal-state properties of high-angle grain boundaries in $(\textmd{Y}, \textmd{Ca}) \textmd{Ba}_2\textmd{Cu}_3\textmd{O}_{7-\delta}$, Phys. Rev. B 71, 094509 (2005).
 \bibitem{lab4}S. H. Christie, The bakerian lecture: Experimental determination of the laws of magneto-electric induction in different masses of the same metal, and of its intensity in different metals, Philos. T. R. Soc. Lond. 123, 95 (1833)
   \bibitem{lab5}C. Wheatstone, Xiii. the bakerian lecture.\&\#x2014; an account
of several new instruments and processes for determining the
constants of a voltaic circuit, Philos. T. R. Soc. Lond. 133, 303
(1843).
   \bibitem{lab6}Giovannetti, S. Lloyd, L. Maccone, Quantum-Enhanced measurements: beating the standard quantum limit, Science \textbf{306}, 1330 (2004).
   \bibitem{lab7}Luca Razzoli, Luca Ghirardi, Ilaria Siloi, Paolo Bordone, and Matteo G. A. Paris, Lattice quantum magnetometry,
Phys. Rev. A 99, 062330 (2019)
    \bibitem{lab8}Kejie Fang, Victor M. Acosta, Charles Santori, Zhihong Huang, Kohei M. Itoh, Hideyuki Watanabe, Shinichi Shikata, and Raymond G. Beausoleil, High-Sensitivity Magnetometry Based on Quantum Beats in Diamond Nitrogen-Vacancy Centers, Phys. Rev. Lett. 110, 130802 (2013).
\bibitem{lab9}Squeezed-Light Optical Magnetometry
Florian Wolfgramm, Alessandro Cer\`{e}, Federica A. Beduini, Ana Predojevi\'{c}, Marco Koschorreck, and Morgan W. Mitchell
Phys. Rev. Lett. 105, 053601 (2010).
\bibitem{lab10}M. T. Mitchison, T. Fogarty, G. Guarnieri, S. Campbell, T.
Busch, and J. Goold, In Situ Thermometry of a Cold Fermi
gas via Dephasing Impurities, Phys. Rev. Lett. 125, 080402
(2020).
\bibitem{lab11}D. Xie, F. Sun, and C. Xu, Quantum thermometry based on a cavity-QED setup, Phys. Rev. A 101, 063844 (2020).
\bibitem{lab12}M. Mehboudi, A. Sanpera, and L. A. Correa, Thermometry
in the quantum regime: Recent theoretical progress, J. Phys.
A: Math. Theor. 52, 303001 (2019).
\bibitem{lab13}S. Campbell, M. Mehboudi, G. D. Chiara, and M.
Paternostro, Global and local thermometry schemes in
coupled quantum systems, New J. Phys. 19, 103003
(2017).
\bibitem{lab14}A. H. Kiilerich, A. De Pasquale, and V. Giovannetti,
Dynamical approach to ancilla-assisted quantum thermometry, Phys. Rev. A 98, 042124 (2018).
\bibitem{lab15}L. A. Correa, M. Mehboudi, G. Adesso, and A. Sanpera,
Individual Quantum Probes for Optimal Thermometry,
Phys. Rev. Lett. 114, 220405 (2015).
\bibitem{lab16}D. Lang, and Carlton M. Caves, Optimal quantum-enhanced interferometry
Matthias,  Phys. Rev. A 90, 025802 (2014).
\bibitem{lab17}Su-Yong Lee, Yong Sup Ihn, and Zaeill Kim, Quantum illumination via quantum-enhanced sensing, Phys. Rev. A 103, 012411 (2021).
\bibitem{lab18}M. Sanz, U. Las Heras, J.J.Garc\'{\i}a-Ripoll, E. Solano, and R. Di Candia, Quantum Estimation Methods for Quantum Illumination, Phys. Rev. Lett. 118, 070803 (2017).
\bibitem{lab19}Si-Hui Tan, Baris I. Erkmen, Vittorio Giovannetti, Saikat Guha, Seth Lloyd, Lorenzo Maccone, Stefano Pirandola, and Jeffrey H. Shapiro, Quantum Illumination with Gaussian States,
Phys. Rev. Lett. 101, 253601 (2008).
\bibitem{lab20}Kasper Poulsen, Alan C. Santos, and Nikolaj T. Zinner,  Quantum Wheatstone Bridge,
Phys. Rev. Lett. 128, 240401 (2022).
\bibitem{lab21}Samuel L. Braunstein, Carlton M. Caves G. J. Milburn, Generalized Uncertainty Relations: Theory, Examples, and Lorentz Invariance, Ann. Phys. (NY) 247, 135 (1996).
\bibitem{lab22}Samuel L. Braunstein and Carlton M. Caves, Statistical distance and the geometry of quantum states, Phys. Rev. Lett. 72, 3439 (1994).
\bibitem{lab23}H. Cram\'{e}r, Mathematical Methods of Statistics (Princeton University Press, Princeton, 1999).
\bibitem{lab24}C. R. Rao, Breakthroughs in Statistics (Springer, New York, 1992).
\bibitem{lab24a}Christian Weedbrook, Stefano Pirandola, Ra¨²l Garc¨ªa-Patr¨®n, Nicolas J. Cerf, Timothy C. Ralph, Jeffrey H. Shapiro, and Seth Lloyd, Rev. Mod. Phys. 84, 621 (2012).
\bibitem{lab25}O. Pinel, P. Jian, N. Treps, C. Fabre, and D. Braun, Quantum parameter estimation using general single-mode Gaussian states, Phys. Rev. A \textbf{88}, 040102(R) (2013).

\bibitem{lab26}Mengzhen Zhang, William Sweeney, Chia Wei Hsu, Lan Yang, A. D. Stone, and Liang Jiang, Quantum Noise Theory of Exceptional Point Amplifying Sensors, Phys. Rev. Lett. 123, 180501 (2019).
 \bibitem{lab27}D. Zhang, X. Q. Luo, Y. P. Wang, T. F. Li, and J. Q. You,
Observation of the exceptional point in cavity magnon polaritons, Nat. Commun. 8, 1368 (2017).
\bibitem{lab28}Y. Sun, W. Tan, H. Q. Li, J. Li, and H. Chen, Experimental Demonstration of a Coherent Perfect Absorber
with PT Phase Transition, Phys. Rev. Lett. 112, 143903 (2014).
 \bibitem{lab29}X.-W. Xu, Y. Li, A.-X. Chen, and Y.-X. Liu, Nonreciprocal
Conversion between Microwave and Optical Photons in
Electro-Optomechanical Systems, Phys. Rev. A 93, 023827
(2016).
 \bibitem{lab30}G. A. Peterson, F. Lecocq, K. Cicak, R. W. Simmonds, J. Aumentado, and J. D. Teufel, Demonstration of Efficient Nonreciprocity in a Microwave Optomechanical Circuit, Phy Review X 7, 031001 (2017).
  \bibitem{lab31}S. Barzanjeh, M. Wulf, M. Peruzzo, M. Kalaee, P.B. Dieterle, O. Painter, and J. M. Fink, Mechanical on-chip microwave circulator, Nat Commun 8, 953 (2017).



\end{thebibliography}
\end{document}